\documentclass[twocolumn]{revtex4}

\usepackage{graphicx}

\begin{document}

\title{
Sociophysics and the Forming of Public Opinion: \\ Threshold versus Non Threshold Dynamics
}
\author{Serge Galam}
\email{serge.galam@polytechnique.edu}
\affiliation{Centre de Recherche en \'Epist\'emologie Appliqu\'ee (CREA),\\
\'Ecole Polytechnique and CNRS, \\1 Rue Descartes, 75005 Paris, France}

\begin{abstract}
A  two state model for opinion forming, which has proven heuristic power,  is reviewed with a novel emphasis on the existence or absence of a threshold for the dynamics. Monitored by repeated small groups discussions floater agents update their opinion according to a local majority rule. A threshold makes the initial supports to  flow towards eitheir one of two opposite attractors with only one single opinion. While odd sizes yield a threshold at fifty percent even sizes, which allow the inclusion of doubt at an opinion tie, produces a threshold shift toward either 0 or 1 giving rise to minority opinion spreading. Considering heterogeneous agents like contrarians and inflexibles  turn the dynamics threshold less beyond sone critical values. One unique attractor at fifty-fifty drives then the dynamics. Inflexibles can generate asymmetry and if one sided they erase the threshold ensuring the associated opinion to eventually gain the all population support. It may shed a new and counter intuitive light on some social aspect of the global warming phenomenon.
\end{abstract}
 
\pacs{02.50.Ey, 05.40.-a, 89.65.-s, 89.75.-k, 05.00.00 Statistical physics, thermodynamics, and nonlinear dynamical systems; 89.20.-a Interdisciplinary applications of physics; 89.75.-k Complex systems}

\maketitle

\newpage

\section{Introduction}

Sociophysics was initiated more than twenty five years ago  \cite{strike}. It stayed confined to a very few researchers for nearly two decades within the hostile environment of the physics community \cite{testimony}.  Only recently it became a main stream of research in physics \cite{book, stauffer-book}.  It has since attracted a great deal of research with an impressive growing number of papers published each year in the main international physical journals. Several international conferences and workshops are being hold regularly every year on the topic. The creation of the Journal of Social Physics is one additional contribution establishing sociophysics as a  flourishing and expanding field of research.

Sociophysics covers numerous topics of social sciences and addresses many different problems including  social networks, language evolution, population dynamics, epidemic spreading, terrorism, voting, and coalition formation.  Among them the study of opinion dynamics has become a main stream of research \cite{mosco-1, chopard-1, sorin, deffuant, sznajd, mino, pajot,  krausse, slanina, redner-2, frank-voter, dynamics,  espagnol,  herrmann, schneider, lambiotte-ausloos,  ausloos-religion, fortunato, pierluigi, mariage, martins}. Moreover public opinion is nowadays a feature of central importance in modern societies making the understanding
of its underlining mechanisms a major challenge. Any progress could have drastic effects on the way to tackle sensitive issues to which the global world is confronted.

Our approach to tackle opinion phenomena  relies on a few simple assumptions, which in turn provide a series of astonishing and powerful results \cite{chopard-1, minority, dynamics, hetero}. In particular it allows to discover that  the dynamics of opinion formation follows some flow, whose direction appear to be determined  by the existence of thresholds in the initial public support for the competing issues. Most models yield such threshold dynamics. Indeed, it was shown that they all belong to one single probabilistic sequential scheme \cite{unify}.

It is worth to notice that in 2005, for the first time, a highly improbable political vote outcome was predicted  using a model of sociophysics \cite{mino, hetero}. Moreover the prediction was made several months ahead of the actual vote against all polls and analyses predictions \cite{lehir}. The model deals with the dynamics of spreading of a minority opinion in public debates using a two sate variable system. It applies to a large spectrum of issues including national votes like the recent French vote, behavior changes like smoking versus non-smoking, support or opposition to a military action like the war in Iraq, rumors like the French hoax about September eleven \cite{rumor}, and reform proposals \cite{reformes}. 

The model uses two state variables to study the forming of a public opinion from a public debate. Agents are floaters who discuss in small groups using a one-person-one-argument principle. At each cycle of discussion  they update their individual opinion according to a local majority rule. The associated
dynamics is driven by  repeated local opinion updates operated trough. Local ties may occur in even group sizes. They are solved in favor of either one opinion according to the common believes of the agents. The resulting opinion forming is found to be a threshold dynamics with a separator $a_{c,r}$ which determines the direction of an opinion flow towards either one of two attractors $a_A$ and $a_B$ at which opinion A and opinion B are respectively the winning majority.  When all agents are floaters the two attractors are single opinionated with $a_A=1$ and $a_B=0$ where only one opinion survives the public debate in the full population. 

For an opinion A initial support $a_t>a_{c,r}$, there exists a number $n$ of successive updates which make the flow to reach the A winning attractor with $a_{t+1}<a_{t+2}<\dots <a_{t+n} \approx a_A$. At opposite $a_t<a_{c,r}$ leads to a decreasing series $a_{t+1}>a_{t+2}>\dots >a_{t+m} \approx a_B$ where the number $m$ is different from $n$. They are both integers which can be calculated. Taking small values they diverge at  $a_{c,r}$. 

For odd size update groups, $a_{c,r}=1/2$. Even sizes allow the inclusion of doubt at an opinion tie in a group. At a doubt, the collective belief is evoked to produce a local bias in favor of either one opinion. Such bias may shift $a_{c,r}$ anywhere between $0$ and $1$ depending on the distribution of both the population collective belief and the local updates group sizes. When $a_{c,r}<<1/2$, the associated dynamics gives rise to the occurrence of minority opinion spreading. 

Focusing on the no tie case for which $a_{c,r}=1/2$ we study the effect of including heterogeneous agents like contrarians \cite{contra-1} and inflexibles \cite{inflex} in addition to floaters. It is  found to have drastic effect on the opinion dynamics. 

Contrarians are agents who deliberately opposes the local majority by shifting to the other opinion, whatever is the majority opinion  \cite{contra-1, contra-2, contra-3, contra-4, contra-5}. At very low densities they create a stable coexistence between a majority and a minority with $a_A \neq 1$ and $a_B \neq 0 $. The threshold being unchanged at $1/2$. However, beyond sone critical values, they turn the dynamics threshold-less  \cite{contra-1}. One unique attractor $a_A =a_B=a_{c,r}=1/2$ drives the dynamics. What ever the initial conditions are, the public debate brings the collective opinion at exactly fifty-fifty. This surprising mechanism of threshold erasing  was used to explain the famous 2000 Bush - Gore presidential election in the US. It was then predicted that fifty-fifty elections were about to occur again and often in voting democracies And it did happened several times since like in Germany, Italy, Mexico  \cite{fifty}. The majority level at which contrarians operate can be global instead of local using results from polls. It gives rise to chaotic behavior around fifty percent  \cite{contra-2}.

Inflexible agents are agents who never shift opinion during small group discussions. They were found to produce similar effects as contrarians but with the novelty of asymmetry since the densities of inflexibles for each opinion are usually not equal. In particular considering one sided inflexibles make the associated opinion certain to gain the all population support. Even if an opinion is supported by only a very low density of inflexibles against a huge majority in favor of the other opinion, the debate will reverse the ratio with eventually the all population aligned along with the inflexibles. 

Accordingly the expected democratic character of a free public debate may turn  onto a dictatorial machine to propagate the opinion of a tiny minority against the initial opinion of the overwhelming majority. It may shed a new and counter intuitive light on the social aspect of the global warming phenomenon.

The rest of the paper is organized as follows. The local majority model with only floater model is presented in the next Section. Update groups are all of the same size $r$.  It is found that the $r=2$ case may exhibit a threshold less dynamics. The combination of various sizes is then investigated. Section 3 deals with heterogeneity of the agents with the possibility of  contrarian behavior. A threshold less dynamics is obtained with a perfectly balanced coexistence of both opinions. One sided inflexible agents  are introduced effects in Section 4. Above a small concentration of one sided inflexibles the dynamics is threshold less making the favored opinion certain to gain the support from the full population through the public debate.  The results  shed a new light on some social aspect of the global warming phenomenon. Last Section contains some discussion about the perspectives to make sociophysics a predictive solid field of science with an  emphasis on both the challenges and the difficulties. Extensions and limits of the approach are discussed. 

\section{The local majority model}

The model consists of a group of $N$ agents undergoing a public debate. Each agent $i=1,\dots ,N$ holds either one opinion A or B, denoted by $a_i=\pm 1 $ respectively. Before the public debate is turned on agents did make a choice individually according to their information and belief. Initial global proportions of both opinions are respectively $a_t$ and $(1- a_t)$. 

The opinion dynamics is driven by a series of repeated cycles of local discussions during which agents update eventually their own opinion. In each update, random groups of $r$ agents are formed. Within each group all agents adopt the opinion which has the local majority. Group size $r$ may vary with $r=1, 2, ...L$. 

However using a local majority rule does not operate in case of a tie in an even size group. Then, a common belief ``inertia principle" is applied to select either one opinion A or B with  respective probability $k$ and $(1-k)$, where $k$ accounts for the collective bias produced by the common believes of the group members.  

A $k$ non integer value accounts for the fact that real societies are divided into disconnected subgroups, which may share different collective believes with not everyone discussing with every one \cite{hetero}. For instance in case of a reform proposal, most cultures share the wisdom stating that in case of a doubt, better to keep the situation unchanged. It is a tip to the status quo, which would put $k=0$ if opinion A corresponds to yes to the reform. But for other issues one social subgroup may have a common belief yielding $k=1$ while another one has an opposite common belief with $k=0$, resulting on average on an effective bias $0<k<1$. 

Accordingly, one cycle of local update leads to new proportions $a_{t+1}$ and $(1-a_{t+1})$ with

\begin{equation}
a_{t+1}= \sum_{m=\frac{r+1}{2}}^{r}  {r \choose m} a_{t}^m  (1-a_{t})^{r-m} \ .
\label{pr-odd} 
\end{equation}
for odd sizes, and for even sizes
\begin{equation}
a_{t+1}= \sum_{m=\frac{r}{2}+1}^{r}  {r \choose m} a_{t}^m  (1-a_{t})^{r-m} +k a_{t}^\frac{r}{2}  (1-a_{t})^\frac{r}{2} \ ,
\label{pr-even} 
\end{equation}
where $ {r \choose m}\equiv \frac{r !}{m ! (r-m) !}$ is a binomial coefficient.

This local majority rule mechanism was first introduced in the eighties to study of bottom-up hierarchical voting models. There, groups of agents designate representatives at a higher hierarchical  level using local majority rule and in case of a tie vote, a status quo inertia.

\begin{figure}
\includegraphics[width=.5\textwidth]{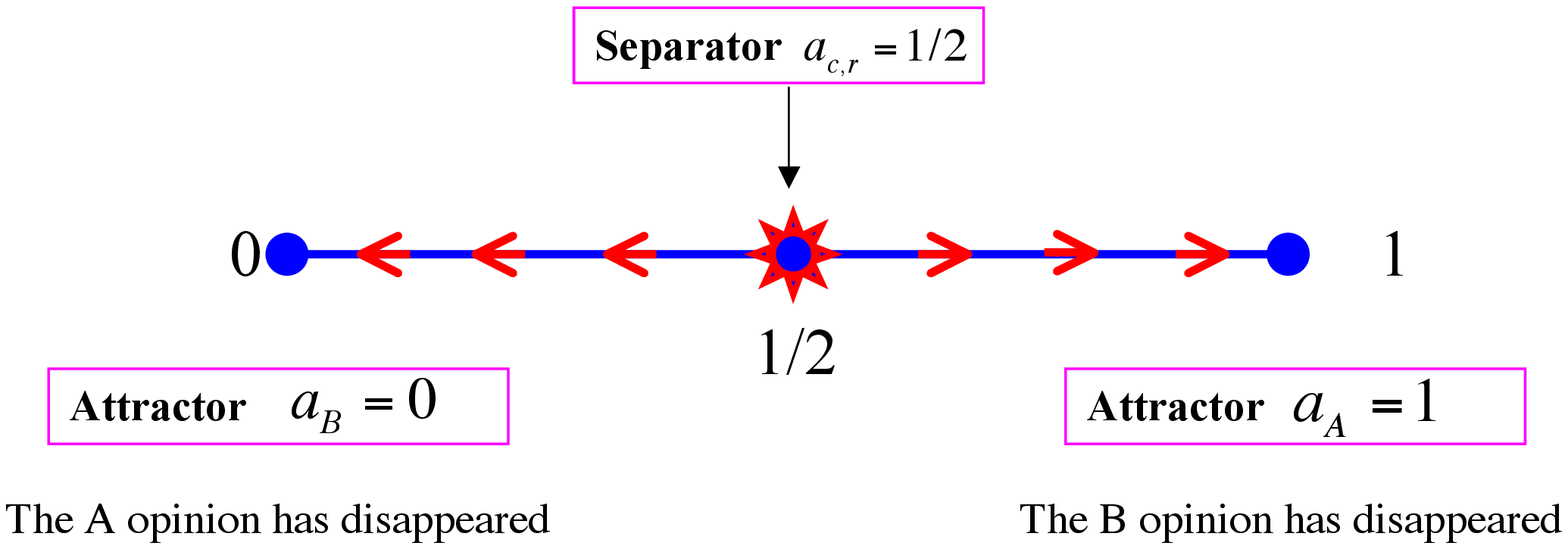}\hfill
\includegraphics[width=.5\textwidth]{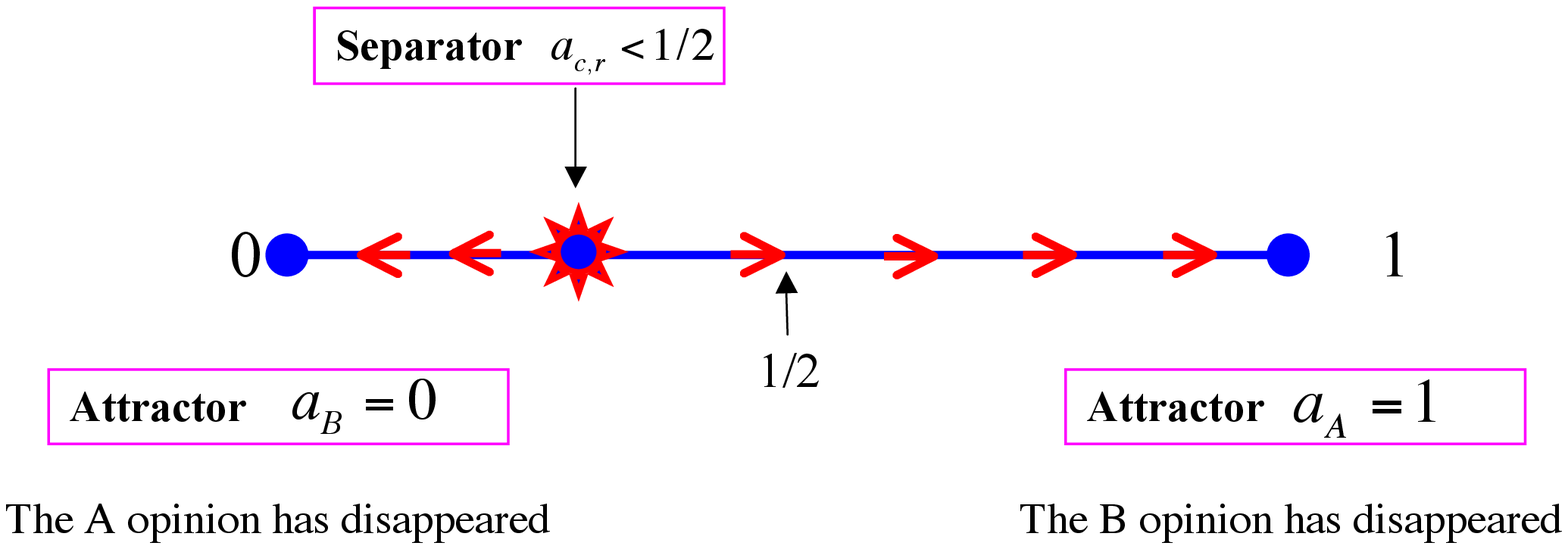}\hfill
\includegraphics[width=.5\textwidth]{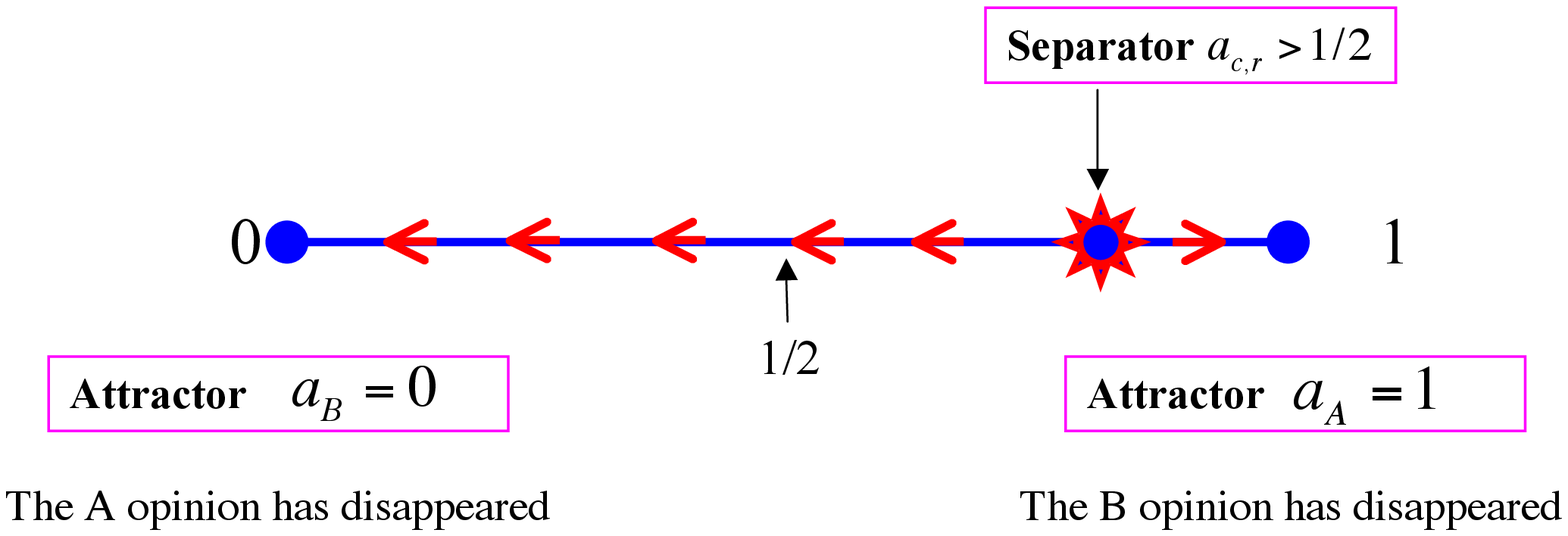}
\caption{The opinion flow for the A opinion with $a_{c,r}=1/2$ (top), $a_{c,r}<1/2$ (middle), and $a_{c,r}>1/2$ (bottom).}
\label{flows}
\end{figure}

To study the dynamics associated with these update Equations we solve the fixed point Equation $a_{t+1}= a_{t}$. It yields two attractors $a_{A}=1$ and $a_{B}=0$ and a  threshold $0<a_{c,r}<1$, which separates the flow opinion in direction of either $a_{A}=1$ or $a_{B}=0$ as shown in Figure \ref{flows} depending on the location of $a_t$ with respect to $a_{c,r}$. if $a_t>a_{c,r}$, one update gives  $a_{t+1}>a_t$. 

Repeating the update gets the support closer to the attractor $a_A=1$. Indeed there exists a number  $n$ to reach it with the series $a_t<a_{t+1}<a_{t+2}< \dots <a_{t+n}=a_{n+1}=a_A=1$ at which an equilibrium state is obtained with only opinion A. The opinion has totally disappear. From $a_t<a_{c,r}$, we get $a_t>a_{t+1}>a_{t+2}>\dots >a_{t+m}=a_{m+1}=a_B=$ where now opinion A has disappeared.

In the case of odd size groups, $a_{c,r}=1/2$. But for even sizes the possibility of local doubt at a tie breaks the symmetry between opinion A and B according to the value of $k$, which is  determined by the distribution of collective believes in the population. The threshold value $a_{c}$ depends on $r$ and $k$ \cite{hetero}. Figure \ref{flows} exhibits the three cases with $a_{c,r}=1/2$ (top), $a_{c,r}<1/2$ (middle), and $a_{c,r}>1/2$ (bottom). For odd sizes and $k=1/2$ for even size $a_{c,r}=1/2$.

Solving the problem exactly in the case of groups of size $r=4$ yields for the threshold
\begin{equation}
a_{c,4}=\frac{6k-1-\sqrt{13-36k+36k^2}}{6(2k-1)} \ ,
\end{equation}
which yields $a_{c,4}=0.23, 0.77, 1/2$ for respectively $k=1, 0, 1/2$. It shows explicitly how the existence of doubt combined with a collective belief, which favors on opinion, here A, can turn the public debate into a machinery to propagate a minority opinion when $k=1$. To win the majority the initial A support must satisfy $a_t>0.23$.

\begin{figure}
\includegraphics[width=.4\textwidth]{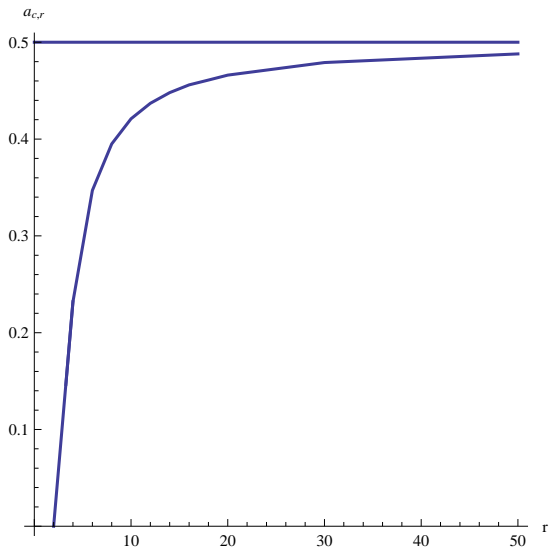}
\caption{The threshold $a_{c,r}$ as a function of even size $r=2,4,6, 8, 10, 12, 14, 16, 18, 20,30,50$ for $k=1$.}
\label{ac-r}
\end{figure}

Increasing the group size $r$ damps this effect by making $a_{c,r}$ closer asymptotically to the value $1/2$. For instance $k=1$ gives respectively $a_{c,r}= 0, 0.23,0.35, 0.39, 0.42, 0.44, 0.45, 0.47, 0.46,0.47$ for $r=2,4,6, 8, 10, 12, 14, 16, 18, 20$ as reported in Figure (\ref{ac-r}). Nevertheless it is worth to emphasize that people discuss in small groups whose size never exceeds a few. Larger groups usually split in smaller ones. 

\subsection{The appearance of a non threshold dynamics}

From Figure (\ref{ac-r}) it appears that while the threshold value decreases with smaller group size, it reaches zero at $r=2$, i.e., for pairwise discussions. It implies a rather strong effect in the opinion dynamics bias driven by  discussions within couples. Indeed many exchanges occur by pairs. It is noticeable that for pairwise groups  the threshold $a_{c,2}=0$, i.e., any A opinion is certain to invade the all population given $k=1$. 

\begin{figure}
\includegraphics[width=.4\textwidth]{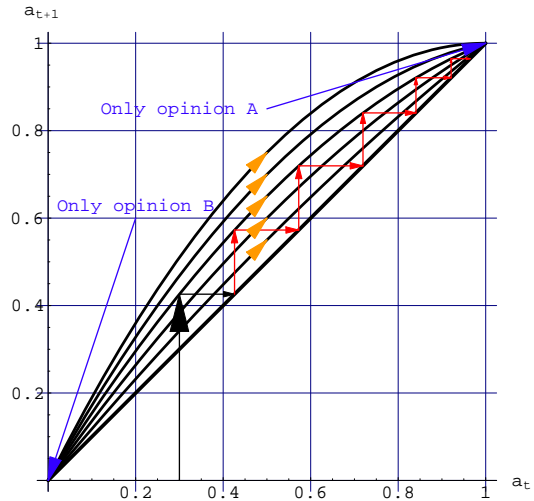}\hfill
\includegraphics[width=.4\textwidth]{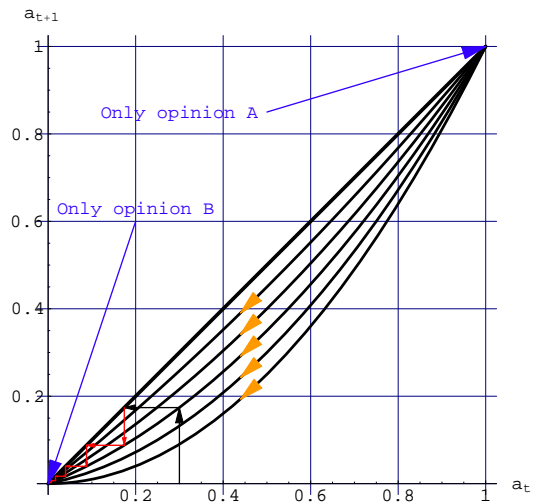}
\caption{The variation of  $a_{t+1}$ as a function of $a_{t}$ from Eq. (\ref{pr-2}). The upper part includes the series $k=1$ (above external line), $k=0.90$ (upper inside), $k=0.80$, $k=0.70$, $k=0.60$, and $k=1/2$ (the diagonal straight line). Any initial A support $a_t\neq 0$ is expanded through the public debate towards total invasion at $a_A=1$.
The lower part includes the series $k=0$ (below external line), $k=0.10$ (below inside), $k=0.20$, $k=0.30$, $k=0.40$, and $k=1/2$ (the diagonal straight line). Any initial A support $a_t\neq 0$ is shrunk through the public debate towards total disappearance at $a_B=0$. The opposite dynamics associated to an initial A support $a_{t}=0.30$ is shown for both cases $k=0.80$ (upper part) and $k=0.20$ (below part).
}
\label{pairs}
\end{figure}

It is the first appearance of a threshold less dynamics. Any initial supports for opinion respectively A and B ends up with the all population sharing opinion A. Unless we start at $a_t=a_{c,2}=0$ with the all population holding opinion B from the start, which in turn means no dynamics at all. Studying the  $r=2$ case more generally Eq. (\ref{pr-even}) writes

\begin{equation}
a_{t+1}= a_{t}^2 + 2 k a_{t}  (1-a_{t}) \ .
\label{pr-2} 
\end{equation}
It yields tow opposite regime, both being threshold less as exhibited in Figure (\ref{pairs}). In the range $0\leq k< 1/2$, the separator is $a_{c,2}=1$ with the unique attractor $a_B=0$. Any initial condition, besides $a_t=1$ leads to $a_{t+n}=0$. At contrast for $ 1/2<k\leq 1$, we have the separator $a_{c,2}=1$ with the unique attractor $a_A=1$. Any initial condition, besides $a_t=0$ leads to $a_{t+n}=1$. The case $k=1/2$ produces  an invariant dynamics with $a_{t+1}=a_{t}$. Agents do change their opinions individually but on averaged the global supports do not.

It is worth to stress that such an effect has been corroborated from a nice analysis of data from the 2004 American presidential election to analyze the marriage gap, i.e. the difference in voting for Bush and Kerry between married and unmarried people. It appear to be possible to interpret this marriage gap in terms our model with a positive value of $k$ and Bush denoted by opinion A \cite{mariage}.

\subsection{Mixing the group sizes}

However, in real life people don't discuss always in successive groups of the same size, therefore to make our model more realistic we consider the distribution given by $p_r$ the probability to have a local group of size $r$ with the constraint,
\begin{equation}
\sum_{r=1}^L p_r=1 ,
\end{equation}
where $r=1, 2, ..., L$  stands 
for respective sizes $1, ..., L$ with $L$ being the larger group  \cite{mino}.  Accordingly the general update Equation writes
\begin{eqnarray}
\label{multisize}
a_{t+1}&=&\sum_{r=1}^L p_r \{\sum_{j=N[\frac{r}{2}+1]}^r C_j^r a_{t}^j
(1-a_t)^{r-j} \\
\nonumber&  & 
+k V(r)C_\frac{r}{2}^r a_t^\frac{r}{2}
(1-a_t)^\frac{r}{2}\},
\label{general}
\end{eqnarray}
where $C_j^r\equiv  \frac{r!}{(r-j)! j!}$, 
$N[\frac{r}{2}+1]\equiv $ Integer Part of  $(\frac{r}{2}+1)$ and
$V(r)\equiv {N[\frac{r}{2}]-N[\frac{r-1}{2} ]}$. It gives $V(r)=1$ for $r$ even and $V(r)=0$ for $r$ odd. 

The occurrence of local ties in even size  groups produce from Eq. (\ref{general}) an asymmetry in the  polarization dynamics towards either one of the two competing opinions with a threshold $a_{c,<r>}$, which may be very unfair for one of the two opinions. We illustrate the process choosing for the size distribution the set $p_1=0.10, p_2=0.25, p_3=0.10, p_4=0.30, p_5=0.10, p_6=0.15$, which yield  $a_{c,<r>}=0.214, 1/2, 0.786$ for respectively $k=1, 1/2, 0$.

\begin{figure}
\includegraphics[width=.4\textwidth]{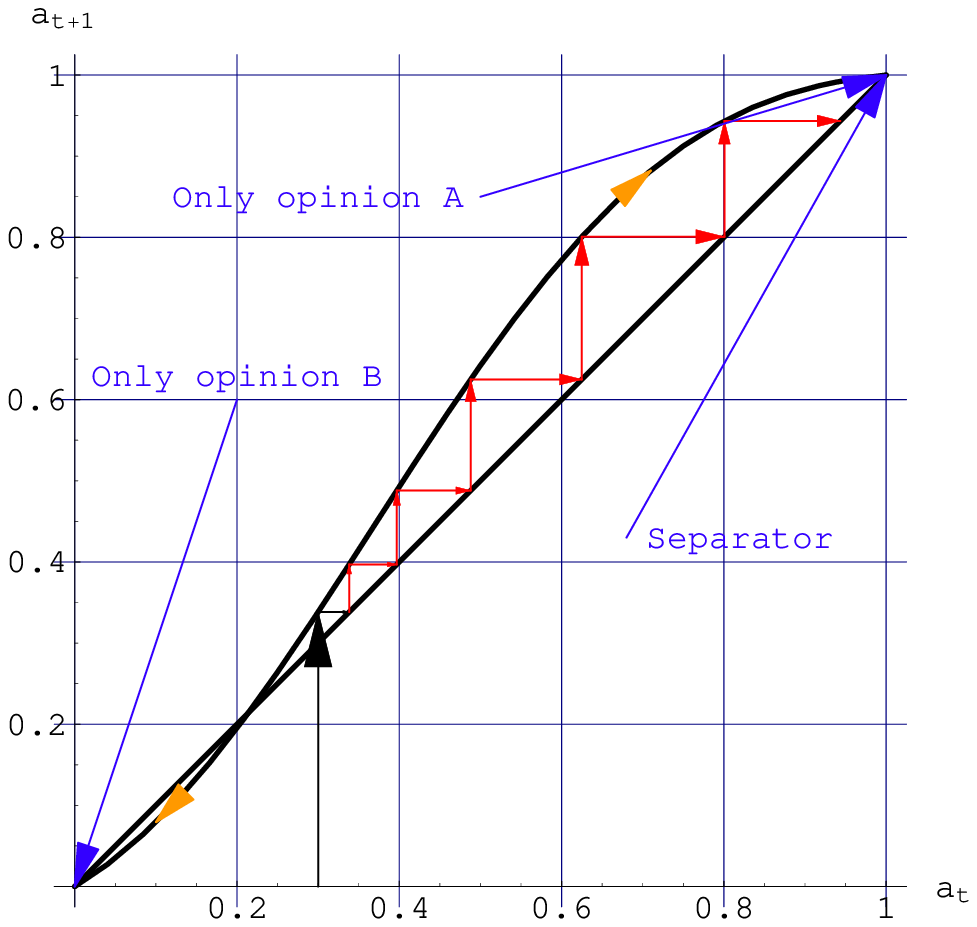}\hfill
\includegraphics[width=.4\textwidth]{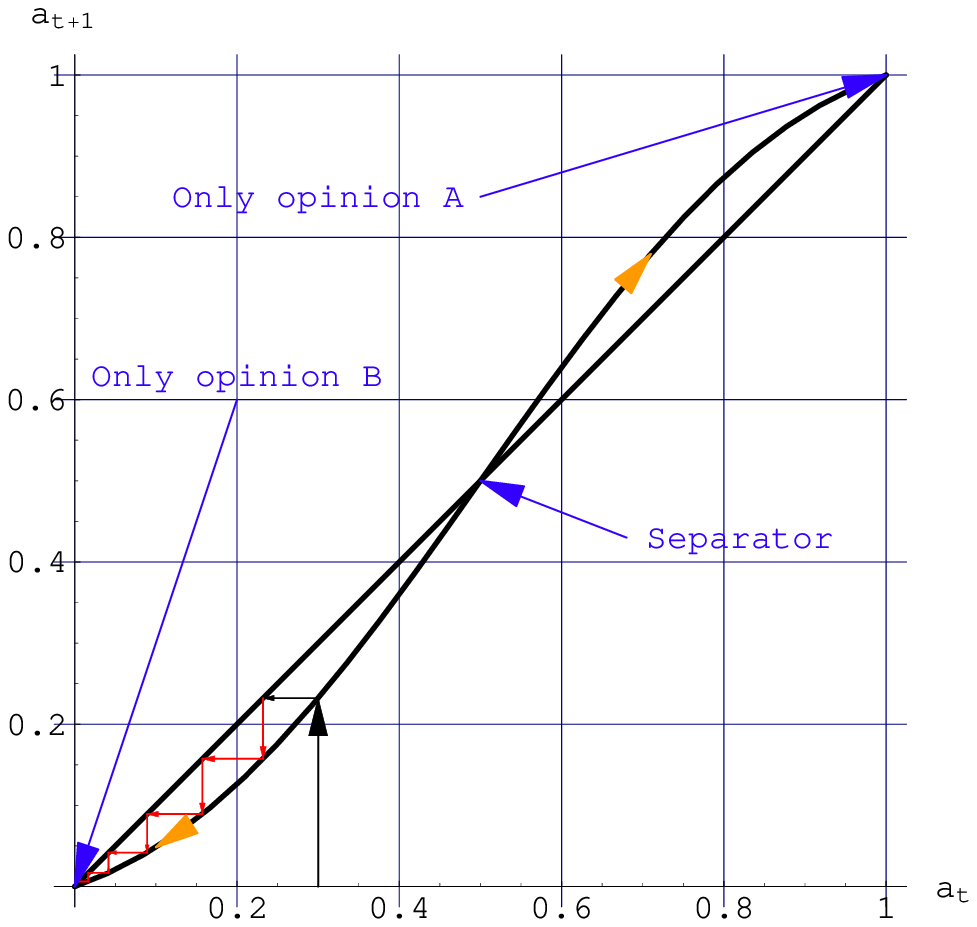}\hfill
\includegraphics[width=.4\textwidth]{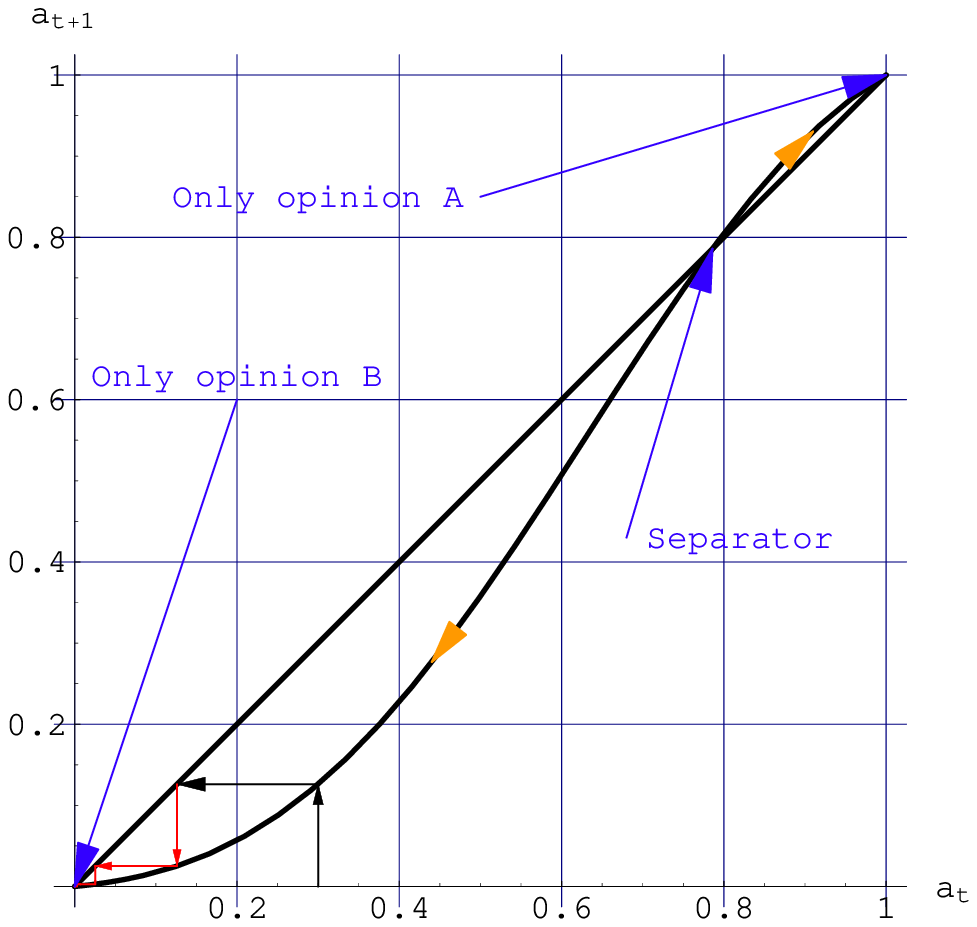}
\caption{ The variation of the general update from Eq. (\ref{general}) for the set of size distribution $p_1=0.10, p_2=0.25, p_3=0.10, p_4=0.30, p_5=0.10, p_6=0.15$ and an initial A support $a_{t}=0.30$. The top part corresponds to $k=1$ which yields $a_{c,<r>}=0.214$. The A minority wins the majority rather quickly. The middle part has a neutral tie with $k=1/2$ which yields $a_{c,<r>}=1/2$. The A support decreases towards zero. The bottom part shows the case of the tie breaking in favor of the B opinion with $k=0$ and $a_{c,<r>}=0.786$. The A support disappears fast.}
\label{flow-k}
\end{figure}

Figure (\ref{flow-k}) illustrates above set of $p_r$ for the three different cases $k=1, 1/2, 0$ for the same initial minority support $a_t=0.30$. When the collective believes favor opinion A, its initial support $a_t=0.30$ is seen to increase drastically driven by the public debate. Only four updates are enough to make opinion A the majority. For a neutral tie effect, it will decrease towards zero while for a bias in favor of opinion B, it falls off very quickly to zero support as seen from Figure (\ref{flow-k}).

\section{Heterogeneous agents and the contrarian effect}

Up to now, we have considered an identical behavior for all agents to make up their opinion. Each one gives its own opinion in a discussing group, and eventually all shift along the same opinion, the one which has received the majority of arguments. In case of a local tie, it is the collective belief bias which sets up the choice. However, it is clear that in real life situations, agents may exhibit different types of behavior in their cognitive process of choosing their opinion. 

Among others, we introduced the contrarian behavior to account for agents who determined their choice, not with respect to some getting convinced criterion but just to oppose what the others think \cite{contra-1, contra-2}. Accordingly a contrarian is someone who deliberately decides to oppose the prevailing choice of the majority around it whatever that choice. 

In the case of update groups of odd sizes  Equation (\ref{pr-odd}) is changed to
\begin{equation}
a_{t+1}=(1-2 c) \sum_{m=\frac{r+1}{2}}^{r}  {r \choose m} a_{t}^m  (1-a_{t})^{r-m} +c\ .
\label{contra-odd} 
\end{equation}
where $c$ is the density of contrarians among the floater population. It shows that the associated dynamics is changed since $a_t=0$ and $a_t=1$ yielding respectively $a_{t+1}=c$ and $a_{t+1}=1-c$, are no longer fixed points of the Equation $a_{t+1}=a_t$. However $a_{c,r}=1/2$ is unchanged as a fixed point.

\begin{figure}
\includegraphics[width=.4\textwidth]{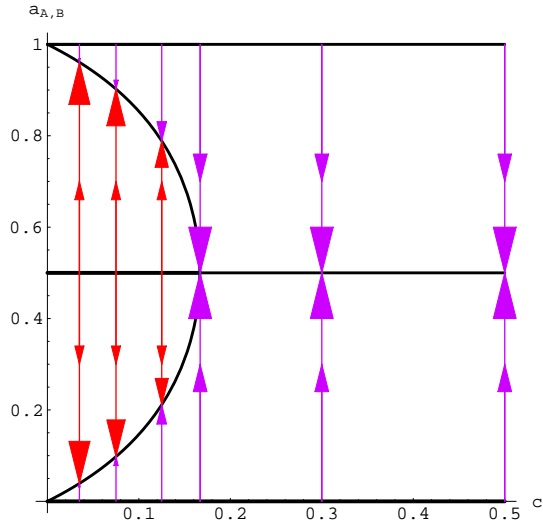}
\caption{ The contrarian two attractors $a_{A,B}^c$  from Eq. (\ref{contra3}) as a function of contrarian density $c$. At $c=1/6)$ both $a_{A}^c$ and $a_{B}^c$ merge with the separator $a_{c,3}=1/2$ to turn the dynamics threshold less. In the range $1/6\leq c\leq  1/2$, the dynamics leads systematically to a fifty-fifty coexistence between opinions A and B.
}
\label{contra-roots}
\end{figure}

Solving the fixed point Equation for $r=3$ yields
\begin{equation}
a_{A,B}^c=\frac{1-2c\pm\sqrt{1-8c+12c^2}}{2(1-2c)},
\label{contra3} 
\end{equation}
which are shown in Figure (\ref{contra-roots}). It is seen that at low densities of contrarians, increasing $c$ makes both attractors to move symmetrically towards the threshold $a_{c,3}=1/2$ with $a_A^c<1$ and $a_B^c>0$. They are found to all merge with $a_A^c=a_B^c=a_{c,3}=1/2$ at exactly $c=1/6$. In the region $0\leq c\leq 1/$ the opinion dynamics yields a stable coexistence between a A majority (B) and a B minority (A). 

\begin{figure}
\includegraphics[width=.5\textwidth]{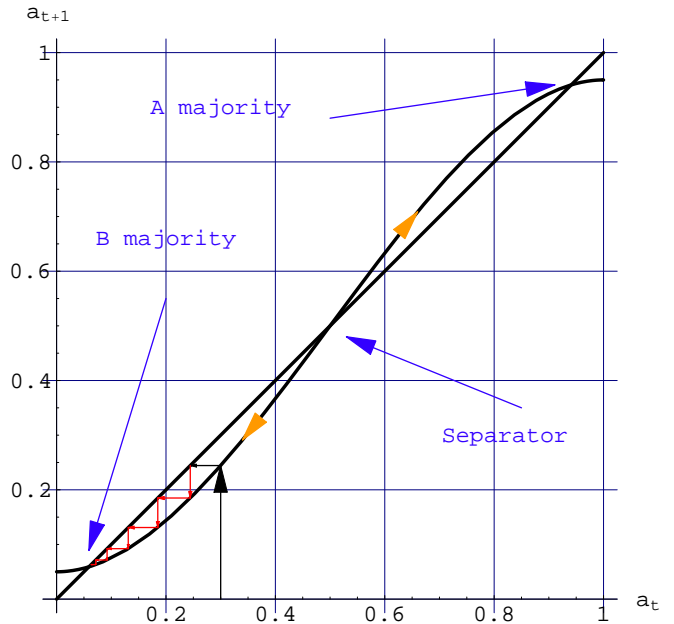}\hfill
\includegraphics[width=.5\textwidth]{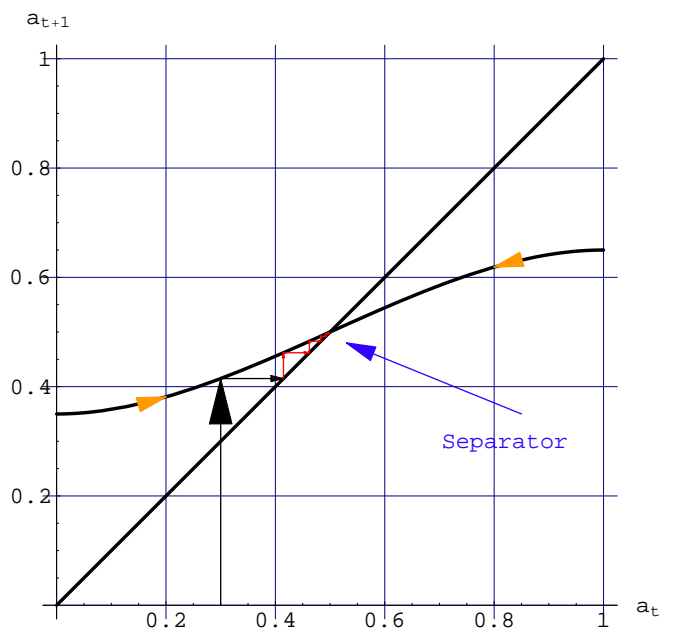}
\caption{The evolution of $a_{t+1}$ as a function of $a_t$ for $c=0.05<1/6$ (upper part) and $c=0.35>1/6$ (bottom part). The evolution of $a_t=0.30$ is shown in both cases. The first one leads to a B majority and a A minority while the second one produces a fifty / fifty support. 
}
\label{contra-small-big}
\end{figure}

Figure (\ref{contra-small-big}) exhibits the two cases of $c=0.05<1/6$ and $c=0.35>1/6$. The evolution of $a_t=0.30$ is shown in both cases. The first one leads to a B majority and a A minority while the second one produces a fifty / fifty support.

\subsection{Threshold less driven coexistence}
From $c=1/6$ and beyond in the range $1/6\leq c\leq  1/2$ the dynamics becomes threshold less. Instead of having a flow which puts ahead one of the two competing opinions, it now reduces any initial difference to zero to establish a perfect equality between the two competing opinions A and B as seen in Figure (\ref{contra-roots}).

Driven by the local discussions and the contrarian effect the two attractors $a_{A}^c$ and $a_{B}^c$  have disappear a the profit of one unique attractor located at  the former separator $a_{c,3}=1/2$. Once the equilibrium is reached, agents keep on updating their opinion due to the contrarians who never settled down on a fixed opinion. However the net effect of these individual opinion changes is perfectly self-balanced. 

Any election to be held in that state yields very narrowed results which are by nature disputable due to incompressible counting errors which score less than the expected statistical fluctuations.

\begin{figure}
\includegraphics[width=.4\textwidth]{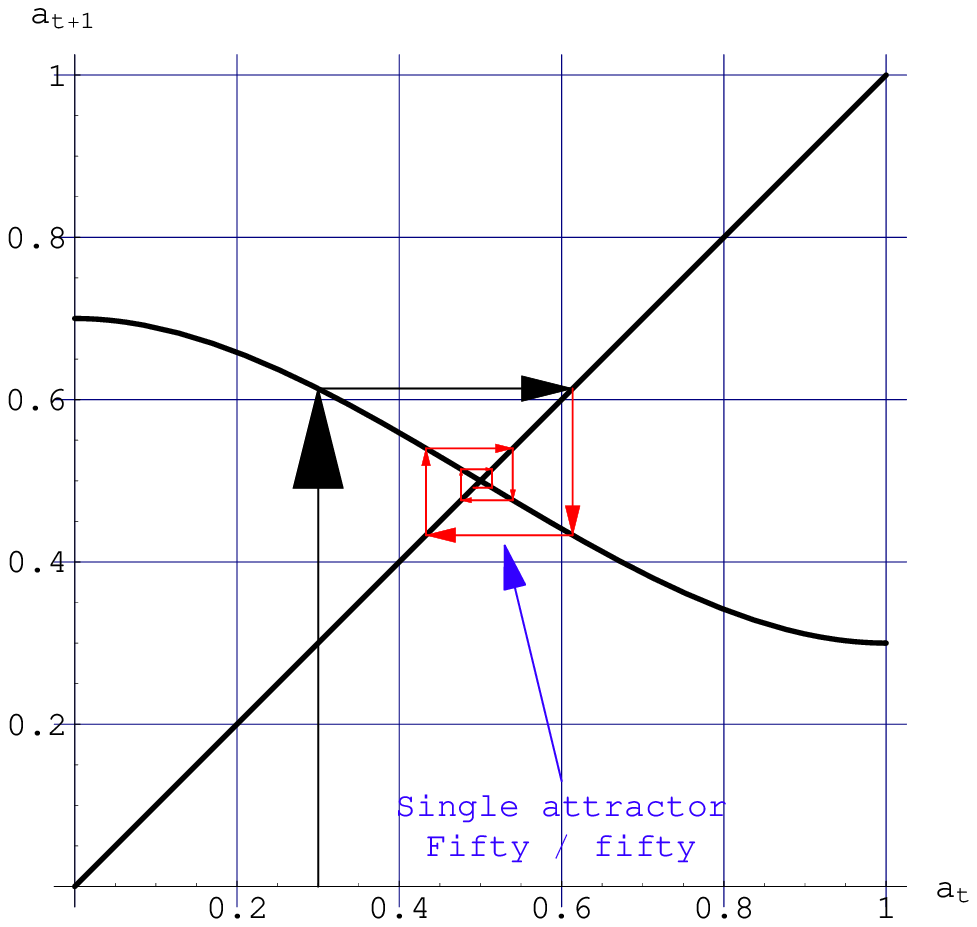}\hfill
\includegraphics[width=.4\textwidth]{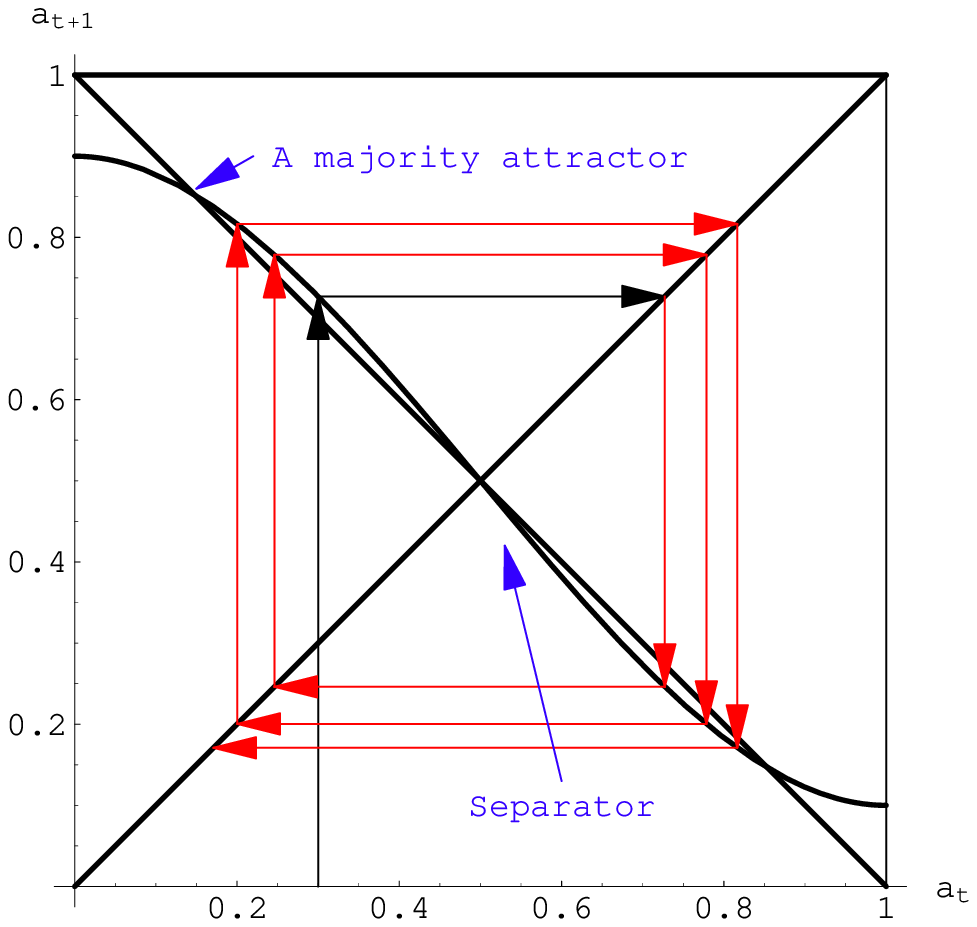}
\caption{The alternating opinion dynamics for $c>1/2$. The top part shows the case $c=0.70$ for which the dynamics is threshold less with a convergence towards the attractor fifty / fifty. The bottom part exhibits the case $c=0.90$ where there exists alternating attractors. The evolution of $a_t=0.30$ is shown for six successive updates in both cases.}
\label{contra-big}
\end{figure}

For larger values of the contrarian density with $c> 1/2$ the dynamics becomes alternating. In the range $1/2\leq c \leq 5/6$, the majority is shifted from one opinion to the other one at each new update, but with the difference in amplitude getting shrunk till reaching zero at the same attractor $1/2$. Beyond with $ 5/6<c\leq 1$, the dynamics is still alternating, but now $1/2$ is again a separator with two alternating attractors $a_A^c<1$ and $a_B^c>0$ given by the fixed point Equation $a_{t+1}=a_t$. Figure (\ref{contra-big}) exhibits the two cases $c=0.70$ and $c=0.90$ for an initial A support $a_t=0.30$.

The contrarian behavior was also extended to the level of the majority at the collective choice given by polls \cite{contra-2}. The effect is similar to the previous one but now a chaotic behavior is obtained around fifty percent.

\section{The one sided inflexible effect and the global warming issue}

Another specific feature of human character is the inflexible attitude. An inflexible agent sticks to its opinion whatever arguments are given to it, it never shift opinion  \cite{inflex}. 

The effect on the dynamics is similar to the contarian effect with the introduction of an asymmetry between the two opinions A and B, depending on the the ratio in the respective proportions of inflexibles. In particular the separator is no longer located at $1/2$ and the two attractors are not any more symmetric  \cite{inflex}.

While it seems natural to have inflexible agents on both sides of any social issue at stake, some peculiar cases do not obey this logic. In particular issues for which some agents are convinced to have a ``scientific proof" to justify their opinion against only doubts prevailing on the other side like for the question of human-caused global warming. 

In this problem, some agents do believe a scientific proof  has been obtained to link the global warming to the man-made production of carbon dioxide. On the other side some agents, much less numerous, claim there is no scientific proof of human culpability. But at the same time they have neither a proof of another cause or the proof men are not guilty. In between these two groups the majority of agents are floaters.

To model the public debate associated to such a situation we consider a proportion $q$ of inflexible agents in favor of opinion A with other agents being floaters who obey a local majority rule. We also assume that initially the majority of floaters support opinion B. For $r=3$ the corresponding update  becomes, instead of Equation (\ref{pr-odd}), 
\begin{equation}
a_{t+1}=a_t^3+3a_t^2 (1-a_t )+q (1-a_t)^2 ,
\label{inflex-q}
\end{equation}
where last term accounts for the configurations where two opinion B holders discuss with one inflexible A agent, yielding the respective opinions unchanged by the local update. It is worth to underline that $a_t\leq q$ by definition of the inflexibles.

To study the effect on the dynamics we solved the new fixed point Equation $a_{t+1}=a_t$, which yields two attractors $a_{A}=1$ and
\begin{equation}
a_{B}^q=\frac{1}{4}\left (1+q-\sqrt {1-6q+q^2} \right) ,
\label{x-p1}
\end{equation}
with a separator located at
\begin{equation}
a_{c,3}^q=\frac{1}{4}\left (1+q+\sqrt {1-6q+q^2} \right) .
\label{ac-inflex}
\end{equation}
Both $a_{B}^q$ and $a_{c,3}^q$ are defined

The first effect of the one-sided inflexibles is to prevent the total disappearance of opinion A even if all the floaters support opinion B. An initial situation with a small inflexible minority $a_t=q$ strengthens the A support to $a_{t+1}=q+q^2-q^3$ after one update. However in the range $0\leq q \leq (3-2\sqrt{2})\approx 0.172$, any A support satisfying $a_t<0.172$ leads to the B opinion to eventually win the public debate. 

On the contrary, any initial support which satisfies $a_t>0.172$ ends ups with a total victory of A opinion even with eighty percent of initial support for opinion B. The process is illustrated in  the upper part of Figure (\ref{inflexes}) with $q=0.05$ and $a_t=0.30$.

\subsection {The threshold less case}

\begin{figure}
\includegraphics[width=.4\textwidth]{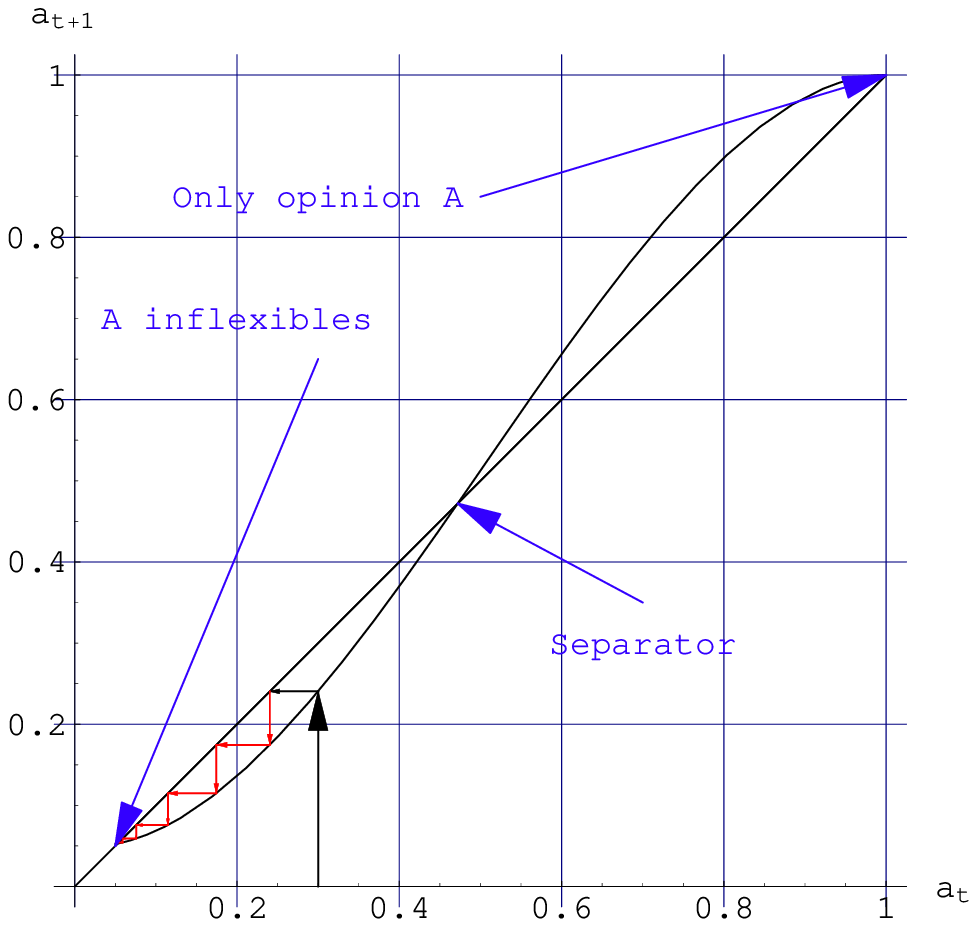}\hfill
\includegraphics[width=.4\textwidth]{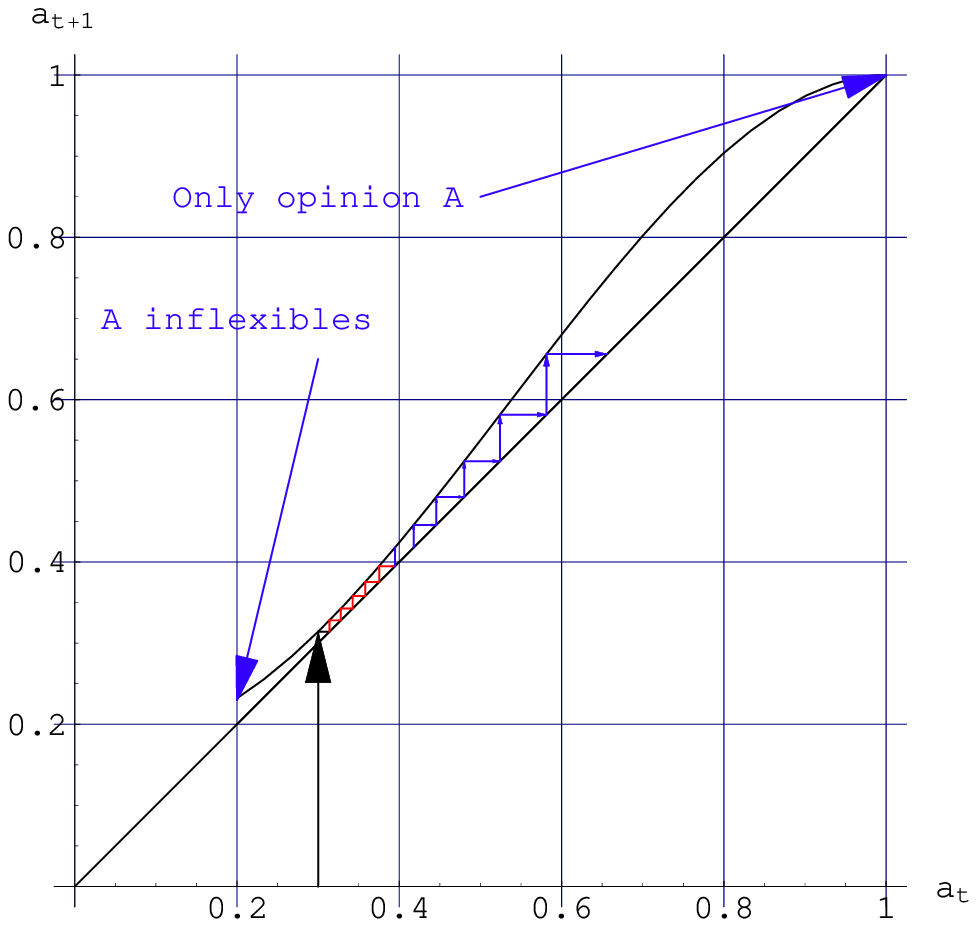}
\caption{The evolution of $a_{t+1}$ as a function of $a_t$ from Eq. (\ref{inflex-q}) for $q=0.05<(3-2\sqrt{2})\approx 0.172$ (upper part) and $q=0.20>(3-2\sqrt{2})\approx 0.172$ (bottom part). The evolution of $a_t=0.30$ is shown in both cases. The first one leads to a very large B majority and a A minority while the second one produces a huge support in favor of opinion A, which eventually gains the full population support. 
}
\label{inflexes}
\end{figure}

The surprising additional effect from including one-sided inflexible agents is that with increasing the inflexible density $q$, both $a_{B}^q$ and $a_{c,3}^q$ move towards one another to eventually merge at $q=3-2\sqrt{2}\approx 0.172$ with $a_{B}^q=a_{c,3}^q\approx 0.293$, and then disappear. One example is exhibited in the bottom part of Figure (\ref{inflexes}) with $q=0.20$ and $a_t=0.30$.

\begin{figure}
\includegraphics[width=.4\textwidth]{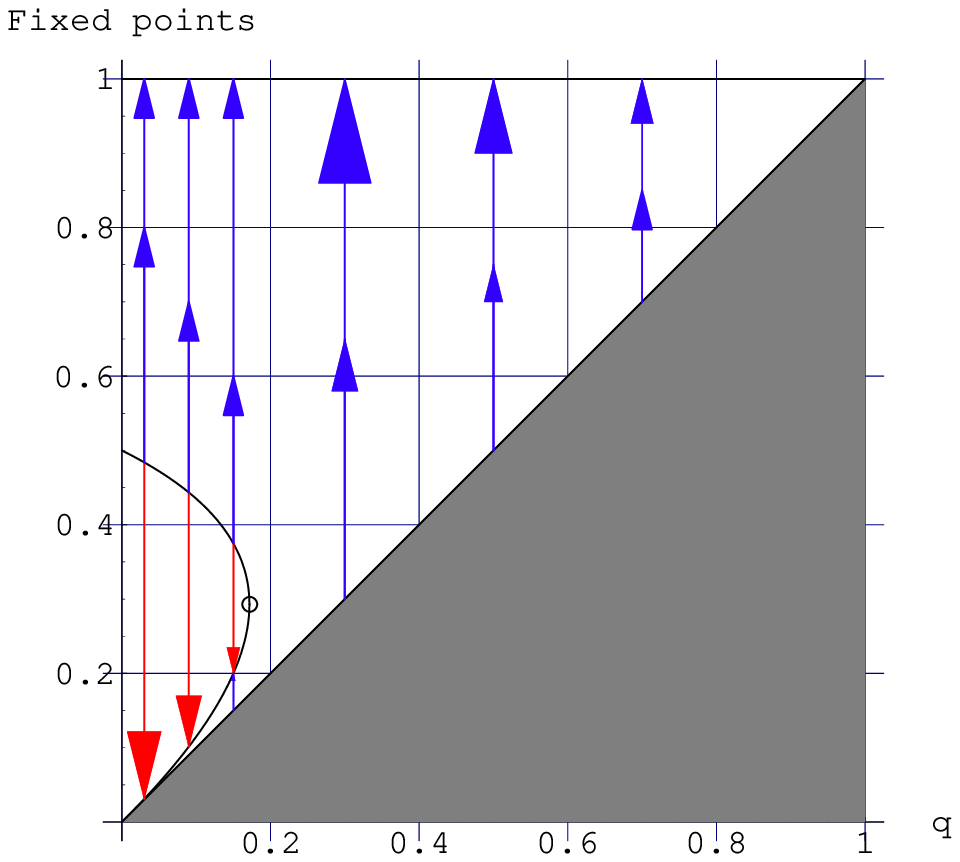}
\caption{The three fixed points from Eq. (\ref{inflex-q})} as a function of the inflexible density $q$. The attractor $a_A=1$ is idendepend of $q$. At opposite both $a_{B}^q$ (lower curve) and $a_{c,3}^q$ (middle curve) gets closer and closer with increasing $q$ to merge and disappear at $q=(3-2\sqrt{2})\approx 0.172$ with $a_{B}^q=a_{c,3}^q\approx 0.293$ (little empty circle). The arrows show the direction of the opinion flow driven by the public deabe. the grey area is the incompressible inlexible proportion of A opinion given by $q$.
\label{inflex-fixedpoints}
\end{figure}

It means that in the range  $q> 0.172$, any B support, even huge ones, systematically looses the public debate against ultra small A minorities. The various regimes are shown in Figure (\ref{inflex-fixedpoints}). The grey area represents the incompressible A inflexible domain. The A attractor $a_A=1$ is independent and stayed unchanged.

Above results may enlighten the mechanisms by which the public debate about the human culpability with respect to the global warming has gained such an increasing support all over the world.  But that is not a formal proof, it is only another way to look at the associated dynamics.

\section{Conclusion}

Sociophysics is a promising field by it specific capacity to reproduce some complex social situations within  a new coherent frame with the discovery of novel and counter intuitive dynamics active in the social reality.

With respect to opinion dynamics it emphasizes the biases which may be  involved in the holding a public debate to decide about some major social  issue. Within our model, we saw  how a free public debate produces quite naturally a dictatorial machine to propagate the opinion of a tiny minority against what could have been the initial opinion of an overwhelming majority. While rationality is applied using a majority rule,  the possible occurrence of local doubts opens the way to have the collective beliefs to make the choice. In case of a threshold dynamics these collective beliefs biases can shift the threshold from fifty percent to any side at values from ten percents to ninety percents.

The model apply  to a large spectrum of social, economical and political phenomena. In particular, propagation effects like fear propagation and rumors spreading. It was used to explain the French hoax about September eleven \cite{rumor}. 

Moreover, in 2005 for the first time, a highly improbable political vote outcome was predicted several months ahead of the actual vote  \cite{lehir, mino}. The victory of the no to the French referendum on the European constitution was confirmed by the actual vote. It is worth to stress that early polls and all analyses were predicting a victory of the yes. The heuristic power of sociophysics was clearly demonstrated as feasible. Nevertheless one result is not sufficient to make definite conclusion. Clearly, more studies are needed.

In this paper we have given some possible explanation to the growing public support for the thesis of human guiltiness  for the current global warming. It was shown why  the associated opinion dynamics will make that opinion of human guiltiness certain to become unanimous without a scientific proof of its rightness. In particular the model shows how powerful is the driving force of individual self-confidence to get a very  small minority to reverse  successfully a huge opposite majority opinion.

To conclude, we have presented a simple model which is able to reproduce some  complexity of the social reality. It suggests that  the direction of the inherent polarization effect in the formation of a public opinion driven by a democratic debate is biased from the existence of common believes within a population. Homogeneous versus heterogeneous situations were shown to result in different qualitative outcomes.

Last, but not least, it is of a crucial importance to keep in mind that we are using models to mimic part of the reality. They are only an approximation of that reality. They are not the reality To forget the difference may lead to some misunderstanding and misleading conclusions of what should be done  the reality with a misuse of the approach. The limits of the approach must be always discussed before making any prediction. At this stage, the collaboration with psycho sociologists as well as political scientists would be welcome.


\end{document}